# Ah-SCDFT: A general approach for superconductivity with anharmonic corrections

Xiaozheng Fan,[1] Panshi Jing,[1] Chuanguang Zhang,[1] Junshuai Wang,[1,*] Chunlan Ma,[2,†] Shijing Gong,[3] Chuanxi Zhao,[4] Tianxing Wang,[1] Yipeng An[1,5,‡]

[1] *School of Physics, Henan Normal University, Xinxiang 453007, China*

[2] *School of Physics and Technology, Suzhou University of Science and Technology, Suzhou 215009, China*

[3] *Department of Physics, East China Normal University, Shanghai 200062, China*

[4] *Guangdong Provincial Key Laboratory of Nanophotonic Manipulation & Department of Physics, Jinan University, Guangzhou 511443, China*

[5] *School of Materials Science and Engineering & Henan Province Engineering Research Center of Flexible Composite and Intelligent Devices, Henan Normal University, Xinxiang 453007, China*

First-principles studies of superconductivity often neglect anharmonic effects (AHE), despite their crucial role in achieving quantitative accuracy in many materials. To bridge this gap, we introduce a general computational approach, termed anharmonic superconducting density functional theory (ah-SCDFT) which systematically incorporates anharmonic corrections into standard SCDFT. This approach allows for high-fidelity predictions of superconducting properties with only a modest increase in computational cost for a limited number of superconducting calculation convergence steps. We demonstrate the effectiveness and reliability of ah-SCDFT by applying it to the prototypical superconductor $MgB_2$, accurately reproducing its superconducting behavior under both ambient conditions and applied pressure in excellent agreement with experiment. Our results establish ah-SCDFT as a powerful, efficient, and broadly applicable approach for quantitatively reliable studies of superconductivity and a promising tool for the prediction of new superconducting materials.

## I. INTRODUCTION

Superconductivity stands as one of the most frontiers in modern physics, embodying both a landmark achievement in fundamental understanding of quantum matter and a cornerstone for transformative technologies. Beyond its historical significance, superconductivity now underpins emerging platforms, such as quantum computing, power transmission and quantum sensing. Superconducting materials are not merely passive constituents but active enablers as they define the coherence properties and performance, including superconducting qubits. Given their dual role, advancing fundamental science while driving technological revolutions, the investigation and prediction of superconducting properties of materials remains a central and enduring pursuit of the global scientific community.

On the theoretical front, a hierarchy of models has been developed to describe superconducting phenomena. Notable approaches include the Bardeen-Cooper-Schrieffer (BCS) theory [1] and its strong-coupling extension, Eliashberg theory [2], as well as anisotropic Migdal-Eliashberg (ME) equations [3], the McMillan/Allen-Dynes semi-empirical formula [4-6], and superconducting density functional theory (SCDFT) [7-10]. Among these approaches, Eliashberg theory is often employed within an isotropic Fermi surface approximation to achieve a practical balance between computational cost and predictive power. In contrast, anisotropic ME equations can capture momentum-dependent superconducting gaps and mode-selective phonon contributions, albeit at substantially increased computational expense. The McMillan/Allen-Dynes formulas provide a convenient estimate of the superconducting critical temperature $T_c$, but their reliance on the empirical Coulomb pseudopotential $\mu^*$, and their neglect of anisotropy limit their reliability in many materials. State-of-the-art SCDFT offers a compelling alternative, combining first-principles rigor with computational efficiency while naturally capturing the anisotropic nature of superconductivity [10,11]. Within this approach, electron-phonon coupling, screened electron-electron repulsion, spin-orbit coupling, plasmonic effects, and spin-fluctuation-mediated interactions can all be treated on equal footing within the harmonic approximation, without requiring empirical parameters such as $\mu^*$.

* Contact author: wangjunshuai@htu.edu.cn

† Contact author: wlxmcl@usts.edu.cn

‡ Contact author: ypan@htu.edu.cn

Nevertheless, it is important to recognize that a wide range of physical phenomena lie beyond the harmonic approximation [12]. For instance, the harmonic treatment breaks down in systems that are dynamically unstable at 0 K, as indicated by imaginary phonon frequencies. It also becomes inadequate in systems with soft degrees of freedom (e.g., molecular crystals) or under conditions involving large atomic displacements from equilibrium, such as those encountered at high temperatures. More broadly, anharmonic effects play a fundamental role in determining the behavior of strongly anharmonic crystals [13-16], governing thermal expansion, lattice thermal conductivity, specific heat, structural phase transitions, phonon linewidths, and the renormalization of electron-phonon coupling (EPC) [17-20]. Notably, several studies have demonstrated that anharmonicity can significantly influence superconductivity in a variety of materials, including 1T-$M_2S$ [17], $CsV_3Sb_5$ [21], $H_3S$ [22], and $KMgH_3$ [23]. These works demonstrate the importance of anharmonic effects in superconductivity research. Nevertheless, AHE are frequently neglected in first-principles simulations of superconductivity. This omission can lead to serious and sometimes qualitative errors, such as: (1) misclassifying a superconducting phase as thermodynamically unstable; (2) misidentifying a system as exhibiting charge density wave; and (3) substantially misestimating the strength of electron-phonon coupling and the resulting superconducting critical temperature $T_c$.

In this paper, we introduce a unified research approach that explicitly integrates anharmonic corrections into the superconducting density functional theory, which we termed "ah-SCDFT". Within this approach, first-principles electronic structure calculations are combined with finite-temperature equilibrium structures obtained from the quasi-harmonic approximation (QHA), enabling a self-consistent determination of the superconducting critical temperature $T_c$. By accounting for temperature-dependent lattice renormalization, ah-SCDFT substantially improves predictive accuracy while incurring only a modest increase in computational cost. We demonstrate the reliability of this approach by accurately reproducing the experimental $T_c$ of $MgB_2$. Moreover, the method remains robust under high external pressure or stress, thereby resolving longstanding discrepancies in theoretical predictions of the pressure dependence of $T_c$ in $MgB_2$ [24,25].

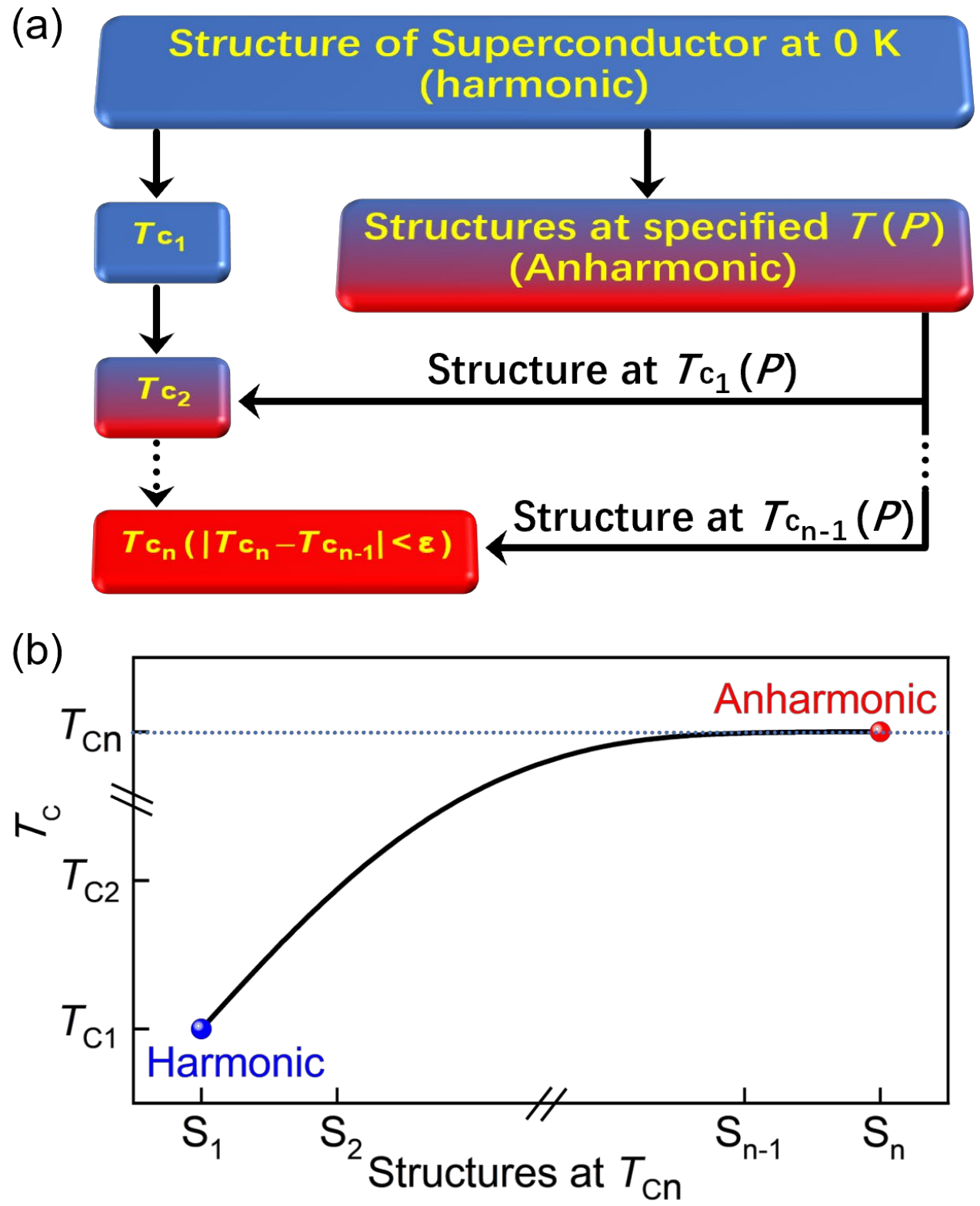

FIG. 1. (a) General approach for studying superconductivity with incorporated anharmonic effects. $T(P)$ denotes the predicted critical temperature $T$ at a given pressure $P$, used for the lattice renormalization. (b) Convergent $T_c$ and $T_{cn}$ under several temperature dependent geometries. The horizontal dashed line marks the final converged $T_{cn}$.

## II. METHODS

Using density functional perturbation theory (DFPT) and SCDFT [8,9,26], we systematically compute the electron and phonon band structures as well as the phonon-mediated superconductivity. All calculations are performed using the QUANTUM ESPRESSO (QE) package [27-29] and SUPERCONDUCTING-TOOLKIT [11,30], adopting the Perdew-Burke-Ernzerhof generalized gradient approximation. In the SCDFT calculations, we employed the Sanna2020 exchange-correlation kernel [10], which combines Eliashberg theory with DFT for accurate predictions of superconducting properties. Note the temperature-dependence equilibrium geometries with the inclusion of anharmonic effects are obtained by the thermo_pw code [31]. Unless stated otherwise, the SG15 optimized norm-conserving Vanderbilt pseudopotentials [32-34] are used to treat core-valence interactions. A plane-wave kinetic energy cutoff of 80 Ry is set for the wave functions, with a corresponding cutoff of 320 Ry for the charge density and potential. Self-consistent calculations are performed using a $12^3$ Monkhorst-Pack **k**-point mesh, while a denser $24^3$ mesh is utilized for the calculations of density of states (DOS) and band structures. Phonon properties are calculated with a $6^3$ **q**-point mesh to balance accuracy and computational cost. The optimized tetrahedron method [35] is employed for Brillouin-zone integration to evaluate the EPC interaction. Structural optimization proceeds until the total energy tolerance and the residual force on each atom are below $10^{-8}$ Ry and $10^{-6}$ Ry Bohr$^{-1}$, respectively. For geometry determination, we employ the PseudoDojo pseudopotentials in the PBE parametrization [36], which show better agreement with experimental lattice constants.

## III. RESULTS AND DISCUSSION

### A. General approach of ah-SCDFT method

The proposed general approach for incorporating anharmonic effects in first principles superconductivity calculations [see Fig. 1(a)] proceeds as follows. First, the ground-state geometry ($S_1$) at 0 K is obtained within the harmonic approximation using standard density functional theory. Superconducting properties, including the superconducting critical temperature $T_{c1}$, are then evaluated using established formalisms such as the anisotropic Migdal-Eliashberg equations [3] or superconducting density functional theory [7-10]. This conventional workflow is often adequate for materials where anharmonic effects are weak, and it remains a practical choice when computational resources are limited. However, for systems with significant anharmonicity, this approach frequently yields quantitatively inaccurate or even qualitatively misleading predictions.

To address this issue, superconducting calculations are subsequently carried out using the temperature-renormalized crystal structure corresponding to the initially predicted critical temperature $T_{c1}$. This procedure yields an updated estimate of the critical temperature $T_{c2}$. By iterating this procedure within the ah-SCDFT approach, the critical temperature converges to a self-consistent value $T_{cn}$ when the error satisfies $|T_{cn} - T_{cn\text{-}1}| < \varepsilon$ (e.g., 0.01 K), which corresponds to the final $T_c$ [see Fig. 1(b)]. Here, the temperature-dependent geometry is obtained by minimizing the Helmholtz free energy, described as follows (neglecting the electronic contribution),

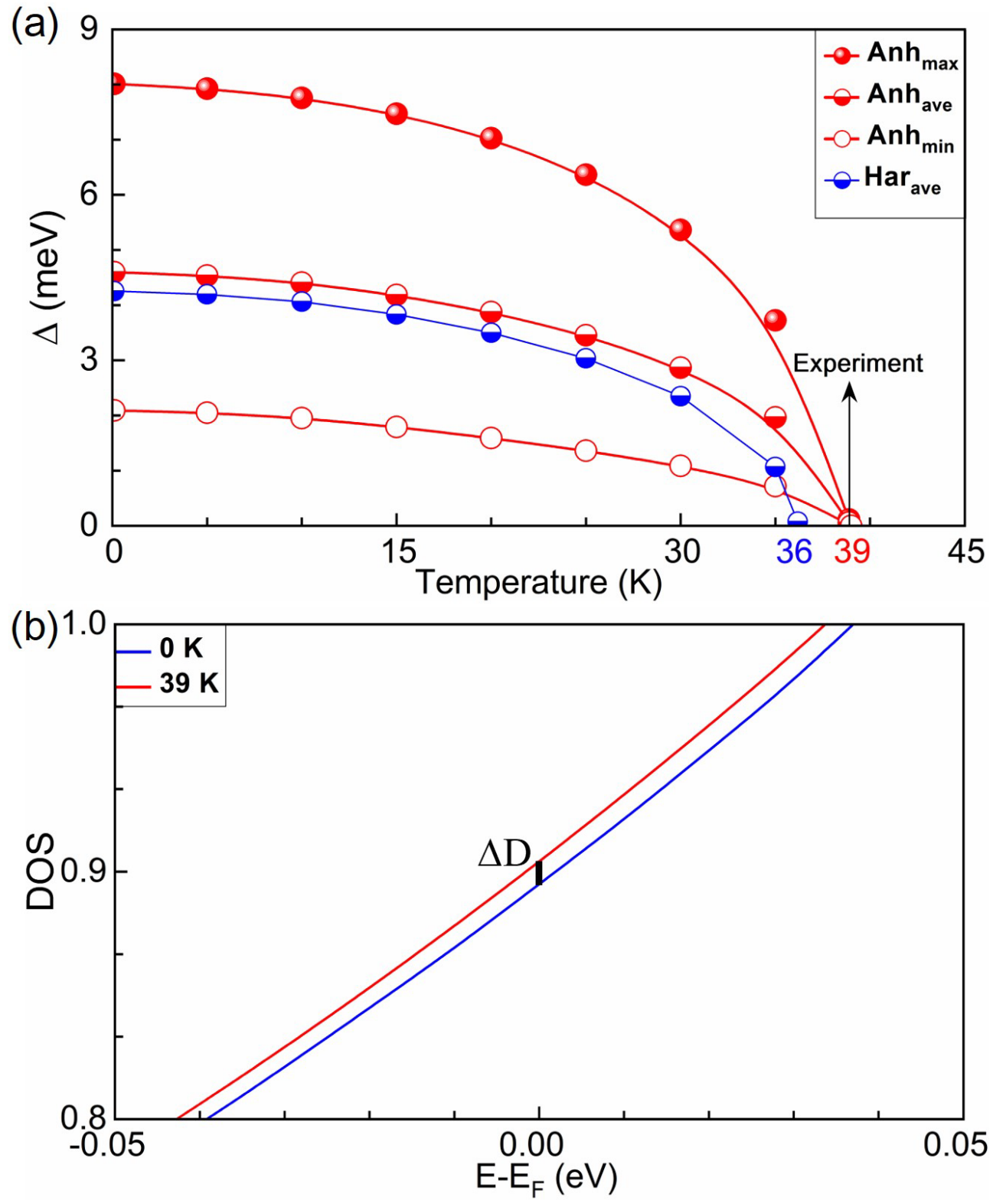

FIG. 2. (a) Temperature-dependent superconducting gaps. The superconducting gaps of $MgB_2$ are shown as a function of temperature, comparing anharmonic (red symbols) and harmonic (blue symbols) calculations. Solid, semi-filled, and hollow red circles represent the maximum, average, and minimum superconducting gaps obtained with anharmonic effects, respectively, across the temperature range from 0.1 K to 39 K ($T_c$). The blue semi-filled circles correspond to the average superconducting gap calculated using the harmonic geometry. (b) Density of states (DOS) of $MgB_2$ calculated using harmonic (blue line, 0 K) and anharmonic (red line, 39 K) equilibrium geometries.

$$F_H = F_0 + F_{vib}, \tag{1}$$

where $F_0$ refers to the total energy, and $F_{vib}$ is the vibrational free energy given by:

$$F_{vib}(\xi,T) = \frac{1}{2N}\sum_{\mathbf{q}\nu}\hbar\omega_\nu(\mathbf{q},\xi) + \frac{k_B T}{N}\sum_{\mathbf{q}\nu}\ln[1-\exp(-\frac{\hbar\omega_\nu(\mathbf{q},\xi)}{k_B T})] \tag{2}$$

where the sums are taken over all phonon modes, each identified by a wave-vector **q** in the first Brillouin zone and a branch index $\nu$. $\xi$ indicates a crystal deformed with a symmetric strain, $T$ the absolute temperature, $N$ the number of unit cells in the solid, $\hbar$ the reduced Planck constant, $\omega_\nu(\mathbf{q},\xi)$ the phonon frequencies, $k_B$ the Boltzmann constant. Moreover, at a specific pressure $P$, the equilibrium structure is found from the minimization of the Gibbs free energy at that pressure.

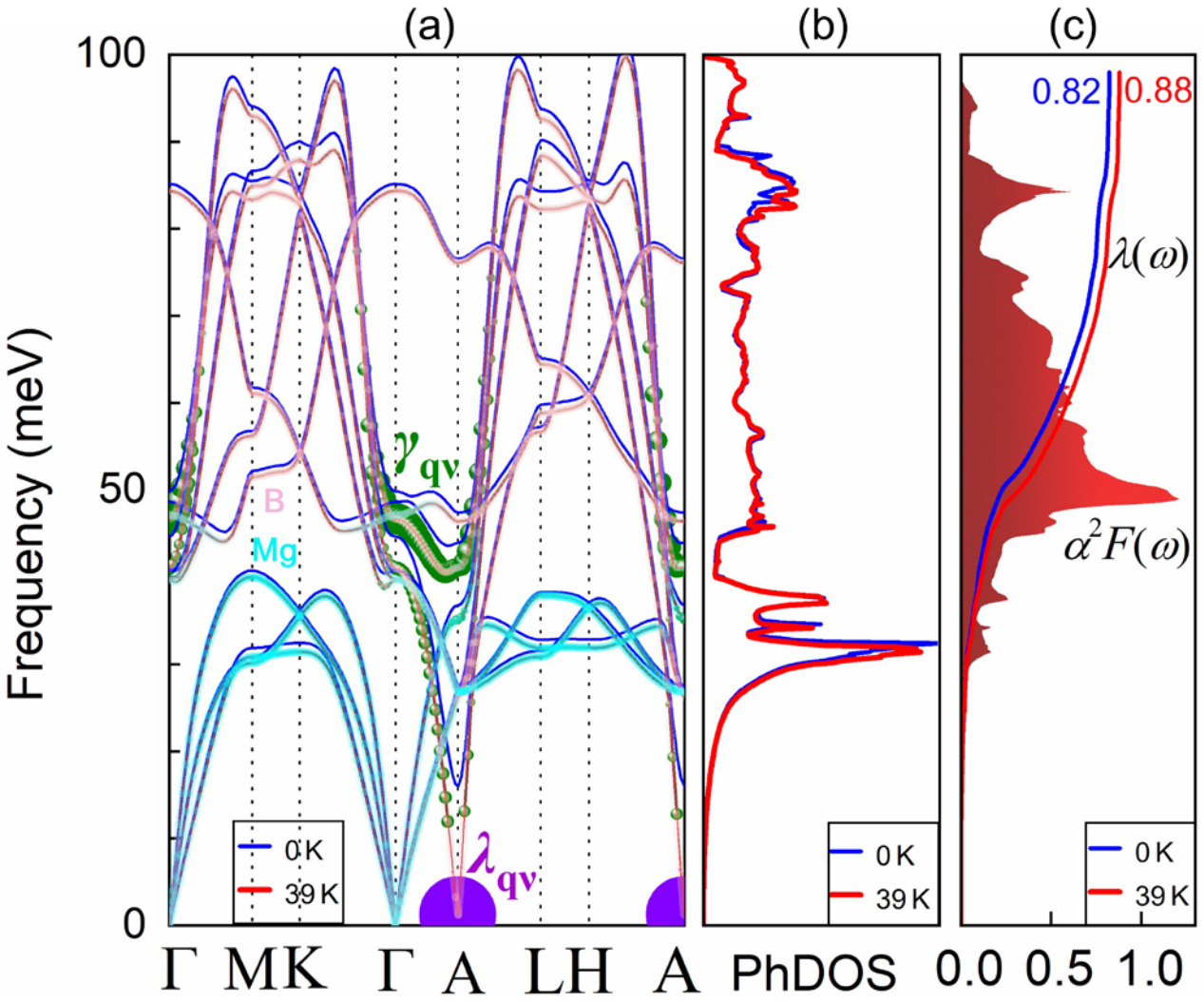

FIG. 3. Phonon and electron-phonon coupling properties of $MgB_2$ within the harmonic approximation (0 K, blue lines) and with the inclusion of anharmonic effects (39 K, red lines). (a) Element-projected phonon dispersion with electron-phonon coupling $\lambda_{\mathbf{q}\nu}$ (purple balls) and phonon linewidth $\gamma_{\mathbf{q}\nu}$ (green balls). (b) Phonon density of states (PhDOS). (c) Frequency-dependent Eliashberg spectral functions $\alpha^2F(\omega)$ with the cumulative EPC strength $\lambda(\omega)$.

### B. Ambient superconductivity of $MgB_2$

To validate the proposed research approach, we select the prototypical $s$-wave superconductor $MgB_2$ as a benchmark system and investigate the effect of anharmonicity on its superconducting properties from first principles. In the calculations for $MgB_2$, the iteration was terminated when $|T_{cn} - T_{cn-1}| < \varepsilon = 0.01$ K, which ensures a well-converged critical temperature. Figure 2(a) presents the temperature dependence of the superconducting gaps obtained within the harmonic approximation (0 K) and with the inclusion of anharmonic effects (39 K equilibrium structure). Within the harmonic approximation, the Fermi-surface-averaged superconducting gap vanishes at 36 K, corresponding to the critical temperature $T_c$ that agrees with the prior reports [10]. Strikingly, when anharmonic lattice renormalization is incorporated through the ah-SCDFT approach, the predicted critical temperature increases to 39 K. At this temperature, the averaged, maximum, and minimum superconducting gaps all close, in excellent agreement with the experimental value [37]. The enhancement of $T_c$ can be traced to an increase in electronic density of states at the Fermi level induced by anharmonic lattice effects in $MgB_2$ structure. Since the DOS at the Fermi level directly enters the electron–phonon coupling constant $\lambda$, a central parameter in the BCS and Eliashberg theories, the anharmonicity-driven increase in DOS, $\Delta D$ ($E_F$) leads to an enhanced $\lambda$ and consequently a higher $T_c$. This correlation between DOS enhancement and superconducting strengthening is illustrated in Fig. 2(b).

To gain deeper insight into the phonon-mediated pairing mechanism and the role of anharmonic effects in $MgB_2$, we next examine its phonon properties. Figure 3(a) presents the element-projected phonon dispersion calculated within the harmonic approximation (0 K) using the 0 K equilibrium structure, as well as the anharmonically renormalized dispersion obtained from the finite-temperature (39 K) lattice. Upon incorporating anharmonic corrections, all phonon branches associated with both Mg and B atoms are somewhat softened across the Brillouin zone, reflecting the temperature-induced renormalization of the lattice dynamics. Notably, this softening is most pronounced near the A point. This region is known to host phonon modes with the strongest electron-phonon coupling, particularly those involving in-plane B vibrations. The enhanced phonon softening near the A point therefore leads to a significant increase in the mode-resolved EPC strength $\lambda_{\mathbf{q}\nu}$.

Due to the averaging over the Brillouin zone, the phonon density of states shown in Fig. 3(b) exhibits only minor changes upon the inclusion of anharmonicity, aside from a downward shift of the first Mg-derived peak. In contrast, the weight of phonon linewidth $\gamma_{\mathbf{q}\nu}$ is significant around 49 meV [see Fig. 3(a)], which gives to a prominent feature in the Eliashberg spectral function $\alpha^2F(\omega)$. This leads to a marked increase in the cumulative EPC strength $\lambda(\omega)$ [see Fig. 3(c)], rising from 0.82 in the harmonic approximation to 0.88 when anharmonic effects are included, directly accounting for the corresponding enhancement of $T_c$ from 36 to 39 K, in excellent agreement with experiment [37]. These results highlight the intimate interplay between anharmonic lattice effects and electron-phonon coupling in $MgB_2$ and, more broadly, demonstrate that anharmonicity must be properly accounted for to achieve reliable predictions of superconducting critical temperatures in real materials.

### C. Superconductivity of $MgB_2$ under external pressures

Applying external pressure is a powerful means for tuning superconductivity, and has enabled the discovery of numerous high-pressure, high-temperature superconductors [38,39]. Despite extensive study, however, the superconducting behavior of $MgB_2$ under high pressure remain a controversial. For instance, Singh *et al*. reported that the critical temperature $T_c$ of $MgB_2$ initially decreases with increasing pressure and then recovers, with a turning point near 100 GPa [40]. In contrast, Ma *et al.* observed a monotonic decrease of $T_c$ up to 200 GPa, beyond which superconductivity vanishes altogether [41]. Other studies have even suggested that the $T_c$ of $MgB_2$ may actually be enhanced under pressure [43]. The pressure-dependent lattice constants including anharmonic effects are shown in Fig. 4(a). Both the lattice constants $a$ and $c$ decrease monotonically with increasing pressure, as expected physically. Our ah-SCDFT results reconcile these seemingly contradictory trends [see Fig. 4(b)]. While the overall trend follows the nonmonotonic behavior reported by Singh *et al.* [40], the explicit inclusion of anharmonic lattice effects shifts the turning point significantly higher pressures, around 200 GPa. This finding, again, highlights the crucial role of anharmonicity in determining the pressure evolution of superconductivity in $MgB_2$ and provides a consistent approach for understanding prior discrepancies. Quantitatively, our calculated values show the best agreement with experimental data (e.g., $T_c$ = 20 K at 25 GPa [42]), in contrast to previous reports [40, 41].

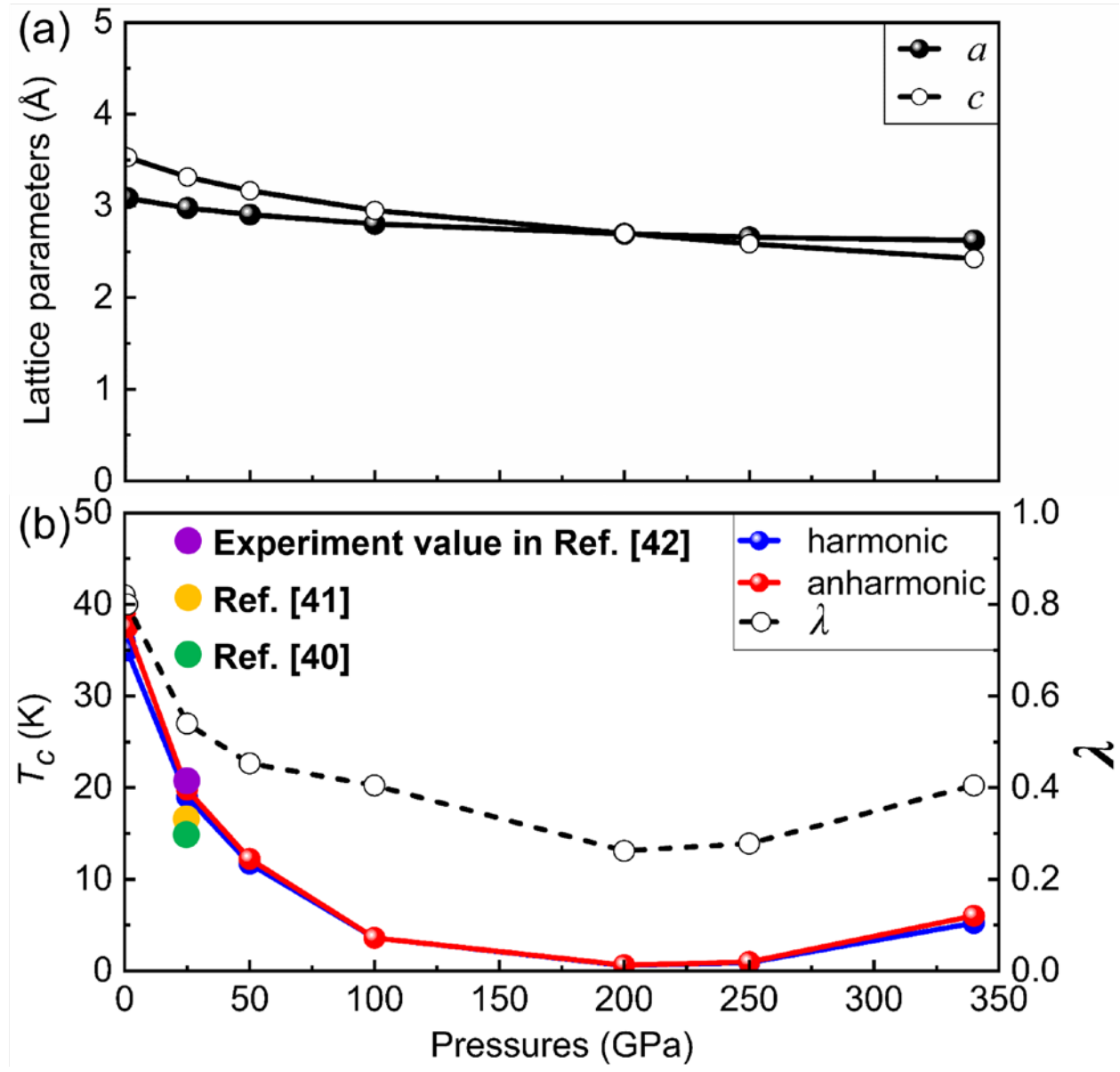

FIG. 4. (a) Pressure-dependent lattice constants including anharmonic effects. (b) Pressure-dependent superconducting critical temperature $T_c$ within the harmonic approximation (blue balls) and with the inclusion of anharmonic effects (red balls), as well as the electron-phonon coupling constant $\lambda$ for the case of including the anharmonic effects.

It is interesting to note that the difference in $T_c$ between the harmonic and anharmonic approximations progressively diminishes with increasing pressure. This behavior is physically intuitive, as external compression stiffens the lattice and thereby suppresses anharmonic effects. Consistent with this trend, the EPC constant $\lambda$ exhibit a pressure dependence analogous to that of $T_c$, as illustrated for the anharmonic case in Fig. 4(b). Finally, $MgB_2$ becomes structurally unstable above 340 GPa, as indicated by the appearance of imaginary phonon frequencies, signaling the onset of a pressure-induced structural phase transition.

## IV. CONCLUSIONS

In summary, we introduce anharmonic superconducting density functional theory, "ah-SCDFT", a general and quantitatively accurate approach for first-principles investigations of superconductivity in real materials. By explicitly incorporating anharmonic lattice effects into SCDFT, this approach enables higher-fidelity predictions of superconducting properties with only a modest increase in computational cost, required by a limited number of superconducting calculation convergence steps. Using the prototypical $s$-wave superconductor $MgB_2$ as a benchmark, we accurately reproduced its superconducting properties under both ambient conditions and applied pressures, achieving excellent agreement with available experimental data. The ah-SCDFT approach thus provides a powerful and broadly applicable tool for the predictive study of phonon-mediated superconductivity and offers a promising route toward the reliable discovery and design of new superconducting materials.

## ACKNOWLEDGMENTS

The work undertaken in China was supported by the National Natural Science Foundation of China (Grant No. 12274117), the Natural Science Foundation of Henan (Grant No. 242300421214 and No. 262300422751), the Program for Innovative Research Team (in Science and Technology) in the University of Henan Province (Grant No. 24IRTSTHN025), the Open Fund of Guangdong Provincial Key Laboratory of Nanophotonic Manipulation (Grant No. 202502), the Guangdong S&T Program (Grant No. 2023B1212010008), and the High Performance Computing Platform of Henan Normal University. We also acknowledge financial support from the National Natural Science Foundation of China (Grants No. 62574079 and No. 12504129).

## DATA AVAILABILITY

The data that support the findings of this article are not publicly available upon publication because it is not technically feasible and/or the cost of preparing, depositing, and hosting the data would be prohibitive within the terms of this research project. The data are available from the authors upon reasonable request.